\documentstyle[preprint,aps,prb]{revtex}

\begin{document}

\newcommand{\xik}{\chi_{\scriptscriptstyle \rm ALDA,\mu\nu}^{-1}}
\newcommand{\fik}{f^{\scriptscriptstyle \rm visc}_{\rm xc,\mu\nu}}
\newcommand{\bfp}{{\bf p}}
\newcommand{\bfk}{{\bf k}}
\newcommand{\bfq}{{\bf q}}
\newcommand{\dy}{\displaystyle}
\newcommand{\sinf}{\raisebox{-.7ex}{$\stackrel{<}{\sim}$}}
\newcommand{\ssup}{\raisebox{-.7ex}{$\stackrel{>}{\sim}$}}
\newcommand{\qp}{q_{||}}
\newcommand{\rp}{r_{||}}
\newcommand{\kp}{k_{||}}
\newcommand{\qqp}{{\bf q}_{||}}
\newcommand{\rrp}{{\bf r}_{||}}
\newcommand{\kkp}{{\bf k}_{||}}
\newcommand{\alda}{^{\scriptscriptstyle \rm ALDA}}
\newcommand{\dlda}{^{\scriptscriptstyle \rm DLDA}}
\newcommand{\vuc}{^{\scriptscriptstyle \rm VUC}}

\ifpreprintsty\else
\twocolumn[\hsize\textwidth%
\columnwidth\hsize\csname@twocolumnfalse\endcsname
\fi
\draft
\tightenlines
\preprint{ }
\title { Collective intersubband transitions in quantum wells: a comparative
density-functional study }
\author {C. A. Ullrich and G. Vignale}
\address
{Department of Physics,  University of Missouri, Columbia, Missouri 65211}
\date{\today}
\maketitle

\begin{abstract}
We use linearized time-dependent (current) density functional theory
to study collective transitions between the two lowest
subbands in GaAs/AlGaAs quantum wells. We focus on two particular systems,
for both of which experimental results are available:
a wide single square well, and a narrow asymmetric double quantum well.
The aim is to calculate the frequency and linewidth of collective 
electronic modes damped through electron-electron interaction only. 
Since Landau damping, i.e.
creation of single electron-hole pairs, is not effective here, the dominant
damping mechanism involves dynamical exchange-correlation effects such 
as multipair production. To capture these effects, one has to go beyond
the widely used adiabatic local density approximation (ALDA) and include
retardation effects.  We perform a comparative study of two approaches 
which fall in this category. The first one is the dynamical extension of 
the ALDA by Gross and Kohn.  The second one is a more recent approach 
which treats exchange and correlation beyond the ALDA as viscoelastic 
stresses in the electron liquid.  It is found that the former method is 
more robust: it performs similarly for 
strongly different degrees of collectivity of the electronic motion.
Results for single and double quantum wells compare reasonably to 
experiment, with a tendency towards overdamping.
The viscoelastic approach, on the other hand, is superior for systems
where the electron dynamics is predominantly collective, but breaks down if the
local velocity field is too rapidly varying, which is the case for 
single-electron-like behavior such as tunneling through a potential barrier.
\end{abstract}
\pacs{71.10.Ca;71.45Gm;73.20Dx,Mf;78.20.Ci;31.15.Ew}
\ifpreprintsty\else\vskip1pc]\fi
\narrowtext


\section{Introduction}

In recent years, time-dependent density functional theory (TDDFT) 
\cite{rungegross,tddft} has become the method of choice for describing 
dynamical electronic properties of atoms, molecules and solids in the 
linear regime and beyond. Until now, TDDFT has been mostly applied within
the adiabatic local density approximation (ALDA)\cite{Zangwill} 
for the dynamical exchange-correlation (xc) potential.  However, some 
propositions have been put forward for improving upon the ALDA. The objective 
of earlier attempts \cite{GK,Dobson} was to obtain  approximations for
the xc potential which would still be local in space, but  not in time.
All these approximations  were found to suffer from inconsistencies,
such as the failure to satisfy the so called ``harmonic potential
theorem'' (HPT) \cite{Dobson,Vignale,Kohn},  and other basic symmetries.
Only recently it has become clear that the root of these difficulties
lies in the fact that the xc potential in TDDFT is an intrinsically nonlocal
functional of the density, that is, a functional that does not admit an
expansion in terms of the density gradients \cite{VK1,VK2}. Consequently,
the most recent extensions of TDDFT beyond the ALDA 
are formulated in terms of the local {\em current} density \cite{VUC}
or the motion of local fluid elements of the electron liquid \cite{buenner}.

Among the variety of phenomena which have been investigated using TDDFT 
methods (see Ref. \onlinecite{tddft} for a recent review),
collective electronic excitations in semiconductor heterostructures   
are currently of great theoretical and experimental interest. In particular,
much effort has gone into the study of parabolic quantum wells with and without
imperfections \cite{parabolic1,parabolic2,parabolic3,bimetallic1,bimetallic2}.
However, the only collective mode that a uniform external field 
can excite in a parabolic well is the well-known center-of-mass mode
\cite{Kohn}, which involves no internal compression of the electron gas.
We therefore focus in this paper on the electronic response in 
single\cite{square1,square2} and double
\cite{double1,double2,double3,double4,double5} 
square wells where no such restriction exists.

The need for going beyond the ALDA becomes evident if one is interested in the
linewidth of these electronic excitations.
The ALDA accounts for Landau damping of collective 
modes, i.e. decay into single particle-hole pairs \cite{pines}. 
For high-frequency, long-wavelength modes, however, this decay
mechanism is not effective, and damping is instead induced by {\em dynamical}
xc effects such as multipair production.
In other words, outside the regime where Landau damping occurs, the modes 
calculated within the ALDA will come out undamped.
This limitation is overcome in non-adiabatic theories. In particular, 
in the formalism of Vignale, Ullrich and Conti (VUC) \cite{VUC}, 
dynamical exchange and correlation 
lead to the appearance of viscoelastic stresses in the electron fluid,
with complex and frequency-dependent viscosity coefficients depending on
properties of the homogeneous electron gas. The viscosity
causes an additional damping not contained in the ALDA.

In an earlier paper \cite{strips}, the VUC formalism was used in a model 
system of two-dimensional quantum strips, where it predicts a substantial
inhomogeneity-induced enhancement of the damping of collective modes
compared to plasmons in the two-dimensional electron gas.
The interest in this model system is based on two reasons: first of all,
it is analytically solvable, and second, it satisfies the conditions of
validity of the approach, namely 
$|\nabla n_0({\bf r})|/n_0({\bf r}) \ll k_F({\bf r})$
and $\omega/v_F({\bf r})$, where $n_0({\bf r})$, $k_F({\bf r})$ and
$v_F({\bf r})$ are the local equilibrium density, Fermi momentum and velocity.
However, no experimental results are available to compare the theoretical
predictions with.

The purpose of the present paper is to compare the performance of various 
non-adiabatic theories in computing the linewidth of
electronic intersubband transitions in single and double square quantum
wells -- systems in which high quality experimental data exist 
\cite{square1,square2,double1,double2,double3,double4,double5}.
Our goal is to clarify the limits of validity of the different approaches,
i.e. the ones by Gross and Kohn \cite{GK} and by VUC (the latter is
equivalent to the method by Dobson {\em et al.} \cite{buenner}
in the situation considered here).
For instance, the formalism of VUC is expected to be
exact in the limit of slowly varying density (see above),
and in particular does satisfy the HPT. However, this does not necessarily
mean that it is the best method under {\em all} conditions: it will 
turn out in this paper that the VUC method works well only for those
situations where the electronic motion is sufficiently collective. 
If, on the other hand, a single-particle-like behavior prevails
(characterized by large gradients in the velocity field), then the
method of Gross and Kohn \cite{GK} gives results which appear to be
in better agreement with experiments.

The paper is organized as follows: In Sec. II we introduce the single and
double quantum wells used in our comparative studies
and calculate their electronic ground state.
Sec. III contains a summary of the linear response formalism and
the various approximations for the linearized xc potential.
In Sec. IV we then calculate the lowest collective intersubband transitions
for the single and double quantum wells and compare the results to experiment.
Conclusions are given in Sec. V.


\section{Electronic ground state of single and double square
quantum wells}

Before discussing the linear response in quantum wells,
we need to find their electronic ground state.
This problem has been dealt with by several authors before
\cite{parabolic2,bimetallic1,double1,double4,dobson1}, and
we give here a summary of the standard procedure to keep this paper
self-contained.

We work in the constant effective-mass approximation, taking 
a value of $m^* = 0.07 m$ for the effective electron mass in GaAlAs.
We furthermore employ a value of 13.0 for the dielectric constant
$\epsilon$, so that the electronic charge becomes $e^* = e/\sqrt{\epsilon}$. 
In the following, we choose units such that $e^* = m^* = \hbar = 1$.
This means that lengths are measured in units of 
1 $a_0^* = 98.3$ \AA, and energies in units of 1 Hartree$^* = 11.27$ meV.  
The difference between the lower conduction band edges of GaAs  
and $\rm Al_{0.3}Ga_{0.7}As$ will be taken as 250 meV.

We assume that the quantum wells have been grown along the $z$-direction
and are translationally invariant in the $x-y$ plane. The Kohn-Sham 
eigenfunctions  for an area $A$ are then discrete in the $z$ 
direction and plane wave-like (with wavevector $\qqp$) perpendicular to it.
They can be written as
\begin{equation} \label{II.A.1}
\psi_{\qqp,j}({\bf r}) = \frac{1}{\sqrt{A}} \: e^{i\qqp \rrp} \varphi_j(z) \;,
\end{equation}
with the Kohn-Sham eigenvalue
\begin{equation} \label{II.A.2}
E(\qp,j) = \frac{\qp^2}{2} + \varepsilon_j \;.
\end{equation}
The orbitals $\varphi_j(z)$ satisfy the following Kohn-Sham equation:
\begin{equation} \label{II.A.3}
\left( -\frac{1}{2} \, \frac{d^2}{dz^2} + v_{\rm ext}(z)
+ v_{\scriptscriptstyle \rm H}(z) + v_{\rm xc}(z)\right)\varphi_j(z) =
 \varepsilon_j  \varphi_j(z) \;.
\end{equation}
Here, the external potential $v_{\rm ext}(z)$
is prescribed by the design of the quantum well.
For the xc potential $v_{\rm xc}(z)$ we use the LDA of
Vosko, Wilk and Nusair \cite{dreizlergross}.
The Hartree potential is obtained by direct integration of Poisson's
equation as \cite{double4} 
\begin{equation} \label{II.A.4}
v_{\scriptscriptstyle \rm H}(z) =
-4\pi \int\limits_{-\infty}^{z} \!dz' \!
 \int\limits_{-\infty}^{z'}\!dz''\: n_0(z'')
+ 2\pi N_s z \;.
\end{equation}
The sheet density $N_s$ is assumed to be a given constant (typically around
$10^{11}\:{\rm cm}^{-2}$). To determine the
density profile in $z$-direction, $n_0(z)$, we write
\begin{eqnarray}\label{II.A.5}
n_0(z) &=&2 \sum_{\qp,j} \varphi_j^2(z)\, \theta(\varepsilon_F - E(\qp,j)) 
\nonumber\\
&=& \frac{1}{2\pi} \sum_{j \atop \varepsilon_j<\varepsilon_F}
\varphi_j^2(z) (q_F^2 - 2\varepsilon_j) \;,
\end{eqnarray}
where $\varepsilon_F$ and $q_F$ are the Fermi energy
and momentum. The sheet density may thus be expressed as
\begin{equation} \label{II.A.7}
N_s = \int\!dz\: n_0(z) = \frac{1}{2\pi} \left( q_F^2 N_{\rm occ}
- 2 \sum_j^{N_{\rm occ}} \varepsilon_j \right) \:,
\end{equation}
where $N_{\rm occ}$ indicates the number of occupied orbitals, and the
orbitals themselves are assumed to be normalized to one.
To close the self-consistency cycle, we now have to determine $N_{\rm occ}$.
To do this, we solve Equation ({\ref{II.A.7}) for $q_F^2$:
\begin{equation}\label{II.A.8}
q_F^2 = \frac{2\pi}{N_{\rm occ}} \: N_s + \frac{2}{N_{\rm occ}}
\sum_j^{N_{\rm occ}} \varepsilon_j \:.
\end{equation}
We now start with $N_{\rm occ} = 1$ and, if necessary,
 keep increasing $N_{\rm occ}$ until the condition
\begin{equation}\label{II.A.9}
\varepsilon_{N_{\rm occ}} < \frac{q_F^2}{2} < \varepsilon_{N_{\rm occ}+1}
\end{equation}
is satisfied. In the cases we are interested in (see below), 
only the lowest subband is occupied, i.e.  $N_{\rm occ} = 1$.

Let us now discuss specific examples of a single and a double square
quantum well, both of which have been experimentally studied.
The single GaAs square well \cite{square1,square2}
is taken to have a width of {384\AA}. Note that this value is still within the
$\pm 4$\% range around the design-width {400\AA} of the sample used in the
experiments \cite{square2}.
The well is filled with electrons up to a sheet density
of $N_s = 0.97\times 10^{11}\:{\rm cm}^{-2}$. In the following, we
consider only the case of zero gate voltage and
neglect any built-in static electric fields.

The (asymmetric)
double quantum well \cite{double1,double2,double3,double4,double5}
consists of two GaAs wells of widths $\rm 85\AA$  and $\rm 73\AA$,
separated by a $\rm 23\AA$ barrier of $\rm Al_{0.3}Ga_{0.7}As$. 
Its electronic sheet density is $N_s = 2\times 10^{11}\:{\rm cm}^{-2}$.
Again, we consider only the case of zero electric field across the sample.

We have solved the Kohn-Sham equation (\ref{II.A.3}) for 
the two systems within the LDA. For our choices of $N_s$, in both cases
only the lowest level is occupied in the ground state. The distribution of
the bound levels is illustrated in Figure 1, which
also indicates the total effective Kohn-Sham potential of the systems.
The energy scale is chosen such that the Fermi level lies at zero.
We see that the single square well supports nine bound levels whose
spacing grows with increasing energy. The lowest
subband spacing is obtained as $E_{12}=8.18$ meV.
The double well, on the
other hand, is found to have only four bound levels.
Here, the subband spacings are
$E_{12}=11.7$ meV, $E_{13}=109.2$ meV and $E_{14}=154.5$ meV.

In the following, we take the ground state of these systems and search for
the lowest collective intersubband transition using linear response theory.

\section{Linear response formalism for quantum wells}

\subsection{Response equation}
The linear density response within TDDFT is given by
\begin{equation}\label{III.A.1}
n_1({\bf r},\omega) = \int\! d^3 r'\: \chi_{\scriptscriptstyle \rm KS}
({\bf r},{\bf r}',\omega) v_{\rm eff}({\bf r}',\omega) \:,
\end{equation}
where
$\chi_{\scriptscriptstyle \rm KS}({\bf r},{\bf r}',\omega)$ is the
noninteracting density-density response function,
which in our case reads
\begin{equation}\label{III.A.2}
\chi_{\scriptscriptstyle \rm KS}({\bf r},{\bf r}',\omega) =
2 \sum_{\mu\nu\atop\kkp,\kkp'}^{\infty} (f_\mu - f_\nu)\:
\frac{\psi_{\kkp,\mu}({\bf r}) \psi_{\kkp',\nu}^*({\bf r})
\psi_{\kkp,\mu}^*({\bf r}') \psi_{\kkp',\nu}({\bf r}')}
{\omega - [E(\kp,\mu)-E(\kp',\nu)] + i\eta} \:,
\end{equation}
with the usual Fermi occupation factors $f_\mu$ and $f_\nu$,
and
\begin{equation}\label{III.A.3}
v_{\rm eff}({\bf r},\omega) = v_{\rm ext,1}({\bf r},\omega)
+ v_{\scriptscriptstyle \rm H,1}({\bf r},\omega)
 + v_{\rm xc,1}({\bf r},\omega) \:.
\end{equation}
Here, $v_{\rm ext,1}({\bf r},\omega)$ is the frequency-dependent external
perturbation, and $v_{\scriptscriptstyle \rm H,1}({\bf r},\omega)$ 
and $v_{\rm xc,1}({\bf r},\omega)$ are
the linearized Hartree and xc potential.  Since our quantum wells
are translationally invariant in the $x-y$ plane, we define
\begin{equation}\label{III.A.4}
n_1(\qqp,z,\omega) = \int\!d^2\rp \: e^{-i\qqp \rrp} n_1({\bf r},\omega)\;,
\end{equation}
so that the response equation (\ref{III.A.1}) may be transformed into
\begin{equation}\label{III.A.5}
n_1(\qqp,z,\omega) = \int\!dz'\:
\chi_{\scriptscriptstyle \rm KS}(\qqp,z,z',\omega) v_{\rm eff}(\qqp,z',\omega)
\;.
\end{equation}
In the following, we shall only be concerned with the case that
$\qp = 0$ in Eq. (\ref{III.A.5}).
The linearized Hartree potential is then
\begin{equation}\label{III.A.6}
v_{\scriptscriptstyle \rm H,1}(z,\omega) =
-4\pi \int\limits_{-\infty}^{z} \!dz' \!
 \int\limits_{-\infty}^{z'}\!dz''\: n_1(z'',\omega) \:.
\end{equation}
The first-order xc potential will be discussed below.

For the Kohn-Sham response function, we follow
Refs. \onlinecite{dobson1,eguiluz} and obtain
\begin{equation} \label{III.A.7}
\chi_{\scriptscriptstyle \rm KS}(\qqp,z,z',\omega) =
\sum_{\mu=1}^{N_{\rm occ}} \sum_{\nu=1}^{\infty} F_{\mu\nu}(\qp,\omega)
\varphi_\mu(z) \varphi_\mu(z') \varphi_\nu (z) \varphi_\nu(z') \:,
\end{equation}
with the definition
\begin{equation} \label{III.A.8}
F_{\mu\nu}(\qp,\omega) = \frac{-2}{(2\pi)^2} \int\!d^2\kp \:
\left\{
\frac{f(\varepsilon_\mu + \kp^2/2)}
{\qqp\kkp + a_{\mu\nu}(\qp) +\omega+i\eta}
+ \frac{f(\varepsilon_\mu + \kp^2/2)}
{\qqp\kkp + a_{\mu\nu}(\qp) -\omega-i\eta} \right\}
\end{equation}
and
\begin{equation}\label{III.A.9}
a_{\mu\nu}(\qp) = \frac{\qp^2}{2} - (\varepsilon_\mu - \varepsilon_\nu) \:.
\end{equation}
Setting $\qp = 0$ in Eqs. ({\ref{III.A.8}) and ({\ref{III.A.9}) 
and performing the integration over $\kp$, we end up with
\begin{equation}\label{III.A.10}
F_{\mu\nu}(\qp=0,\omega)  = 
-\frac{\varepsilon_F-\varepsilon_\mu}{\pi} \left\{
\frac{1} {\varepsilon_\nu - \varepsilon_\mu  +\omega+i\eta}
+ \frac{1} {\varepsilon_\nu - \varepsilon_\mu -\omega-i\eta} \right\} \;.
\end{equation}
This, together with Eq. (\ref{III.A.7}), defines the response
function $\chi_{\scriptscriptstyle \rm KS}(\qqp=0,z,z',\omega)$.
In the following, we will evaluate $\chi_{\scriptscriptstyle \rm KS}$
summing only over the {\em bound} states $\varphi_\nu$, i.e. the ones shown
in Figure 1. This approximation is
justified by the fact that the collective excitations we are studying
are well below the continuum threshold for the quantum wells under 
consideration and thus involve only the lowest-lying states.

\subsection{Approximations for the xc potential}

In our case, the general expression for
$v_{\rm xc,1}$ depends only on the $z$ coordinate, and is given by
\begin{equation} \label{III.B.1}
v_{\rm xc,1}(z,\omega) = \int \! dz'\: f_{\rm xc}(z,z',\omega)\:
n_1(z',\omega)\;,
\end{equation}
where the xc kernel $f_{\rm xc}(z,z',\omega)$ is a functional of the
ground-state density $n_0(z)$.

The simplest possible approximation for $v_{\rm xc,1}$ is the 
ALDA, which assumes
\begin{equation} \label{III.B.2}
f_{\rm xc}\alda(z,z',\omega) =
\delta(z-z')\: \left. \frac{d^2 e_{\rm xc}^{hom}(n)}{dn^2}\right|
_{n=n_0(z)} \;,
\end{equation}
where $e_{\rm xc}^{hom}(n)$ is the xc energy density of the
homogeneous electron gas, so that
\begin{equation} \label{III.B.3}
v_{\rm xc,1}\alda(z,\omega) = n_1(z,\omega)
\: \left. \frac{d^2 e_{\rm xc}^{hom}(n)}{dn^2}\right|_{n=n_0(z)} \;.
\end{equation}
In other words, the xc kernel in the ALDA has no frequency dependence
at all and is purely real. 
This approximation is justified for cases where the external
potential is slowly varying in time as well as in space. 

In a first step towards overcoming the restriction to slow variation
in time, Gross and Kohn \cite{GK} proposed the following
plausible approximation (which we denote by  DLDA, the D standing
for ``dynamic''), 
valid for the case where both $n_0$ and $n_1$
are slowly varying in space:
\begin{equation} \label{III.B.4}
v_{\rm xc,1}\dlda(z,\omega) =  f_{\rm xc}^{hom}
(n_0(z),q=0,\omega)\: n_1(z,\omega)\;,
\end{equation}
where $f_{\rm xc}^{hom}$ is a property of the homogeneous electron gas,
related to the dynamical local field factor as
$f_{\rm xc}^{hom}(n_0,\omega) = -\lim_{q\to 0} 4 \pi G(q,\omega)/q^2$.
Gross {\em et al.}  \cite{GK} have given a simple parametrization
of $f_{\rm xc}^{hom}(n_0,\omega)$ (see also Ref. \onlinecite{tddft}). 
A more elaborate calculation of $f_{\rm xc}^{hom}(n_0,\omega)$ 
has recently been performed by Nifos\`\i, Conti and Tosi \cite{conti,nifosi} 
using an approximate decoupling
scheme of the equation of motion for the current density, which accounts for
processes of excitation of two electron-hole pairs.

As already discussed in the introduction, the DLDA (\ref{III.B.4})
has recently been found to violate the HPT (for
more details see Refs. \onlinecite{VK1,VK2,VUC}).
However, a local frequency-dependent approximation which satisfies the HPT
can be derived if the current density is taken as basic variable rather
than the particle density. The resulting first-order xc vector potential 
in the form stated by VUC \cite{VUC}
is given by
\begin{eqnarray} \label{III.B.5}
i\omega\:a\vuc_{{\rm xc},1,\mu}({\bf r},\omega) &=&
\nabla_{\mu} v_{\rm xc,1}\alda({\bf r},\omega)
\nonumber\\
&-&\frac{1}{n_0({\bf r})} \sum_{\nu} \nabla_\nu
\sigma_{{\rm xc},\mu\nu}({\bf r},\omega) \:.
\end{eqnarray}
The dynamical correction to the ALDA xc potential
is the divergence of the viscoelastic stress tensor
\begin{eqnarray} \label{III.B.6}
\sigma_{{\rm xc},\mu\nu} &=& \eta_{\rm xc}
\Big( \nabla_\nu u_{\mu} + \nabla_\mu u_{\nu}
-\frac{2}{3}\: \nabla \cdot {\bf u} \, \delta_{\mu\nu} \Big)
\nonumber\\
&& {}+\zeta_{\rm xc} \nabla \cdot {\bf u} \, \delta_{\mu\nu} \:.
\end{eqnarray}
Here, ${\bf u}({\bf r},\omega) = {\bf j}_1({\bf r},\omega)/n_0({\bf r})$
is the velocity field, and $\eta_{\rm xc}(\omega,n_0({\bf r}))$ and
$\zeta_{\rm xc}(\omega,n_0({\bf r}))$ are complex viscosity coefficients
whose explicit form is given in Ref. \onlinecite{VUC}.

In the one-dimensional case, it is straightforward to derive the 
associated xc potential \cite{VK2}. It can be written as
\begin{eqnarray}\label{III.B.7}
v_{\rm xc,1}\vuc(z,\omega) 
&=& f_{\rm xc}^{hom}(n_0(z),\omega)\: n_1(z,\omega) \nonumber\\
&&{}- \frac{n'_0(z)}{n_0(z)}\: f_{\rm xc}^{dyn}(z,\omega)
\int\limits _{-\infty}^{\dy z} dz'' \:n_1(z'',\omega)
\:-\: \int\limits _{\dy z}^{\infty} dz' \: \frac{n'_0(z')}{n_0(z')}
\: f_{\rm xc}^{dyn}(z',\omega)\: n_1(z',\omega) \nonumber\\
&&{}+\int\limits _{\dy z}^{\infty} dz' \: 
\left(\frac{n'_0(z')}{n_0(z')}\right)^2 f_{\rm xc}^{dyn}(z',\omega) 
\int\limits _{-\infty}^{\dy z'} dz'' \:n_1(z'',\omega) \:.
\end{eqnarray}
Here, we have defined
\begin{equation}\label{III.B.8}
f_{\rm xc}^{dyn}(z,\omega) \equiv
f_{\rm xc}^{hom}(n_0(z),\omega) -\
\left.\frac{d^2e_{\rm xc}^{hom}(n)}{dn^2}\right|_{n=n_0(z)} \;.
\end{equation}
The first term on the right-hand side of Eq. (\ref{III.B.7}) is the
DLDA (\ref{III.B.4}). The other terms are needed to
satisfy both the HPT as well as Onsager's reciprocity theorem,
i.e. symmetry under interchange of $z$ and $z'$ of the associated
xc kernel $f_{\rm xc}(z,z',\omega)$ (see Ref. \onlinecite{VK2}).

Both the DLDA and the VUC approach lead to a complex xc potential. This 
means that the frequency of collective modes will come out complex as well, 
with a negative imaginary part accounting for the damping. As a consequence,
care has to be taken to perform the proper analytic continuation of the
Kohn-Sham response function and of $f_{\rm xc}$, both of which are analytic
functions in the upper half complex plane, into the lower half complex plane
\cite{pines,dobsonharris1,dobsonharris2}.

The analytic continuation of the response function 
(\ref{III.A.7}) with (\ref{III.A.10}) is straightforward.
This is because there are no branch cuts in the frequency region
we consider, and the depolarization shift will move the mode frequency 
sufficiently far away from the poles in $F_{\rm \mu\nu}(\qp=0,\omega)$.
Thus, $\chi_{\scriptscriptstyle \rm KS}(\qp=0,z,z',\omega)$
is analytic across the real axis in the frequency region of interest.

For the analytic continuation of $f_{\rm xc}$, we proceed as shown
Ref. \onlinecite{dobsonharris2}. The idea is the following: 
let us assume we have a parametrization of  the imaginary part of
$f_{\rm xc}(\omega)$ on the real frequency axis. Denote this parametrization
by $A_{\rm xc}(\nu)$, where $\nu$ is real. Then, the
expression for $f_{\rm xc}(\omega)$ in the {\em upper} complex 
$\omega$ plane is
\begin{equation}\label{III.B.9}
f_{\rm xc}^{\rm up}(\omega) = f_{\infty} +
\frac{1}{\pi} \int\limits_{-\infty}^{\infty}
\frac{A_{\rm xc}(\nu)}{\nu-\omega} \: d\nu \;, \quad \Im \omega > 0 \;.
\end{equation}
Here, $ f_{\infty}$ is the large-frequency limit of $f_{\rm xc}$.
The analytic continuation into the {\em lower} complex $\omega$ plane is
given by
\begin{equation}\label{III.B.10}
f_{\rm xc}^{\rm lo}(\omega) = f_{\infty} +
\frac{1}{\pi} \int\limits_{-\infty}^{\infty}
\frac{A_{\rm xc}(\nu)}{\nu-\omega} \: d\nu 
+ 2i A_{\rm xc}(\omega)
\;, \quad \Im \omega < 0 \;.
\end{equation}
Note that the third term on the right-hand side implies that
$A_{\rm xc}$ has to be evaluated at the complex frequency 
$\omega$, $\Im \omega <0$. Therefore, in moving down from the
real $\omega$ axis into the lower half complex plane, one has to make 
sure along the way that no branch cuts or poles of $A_{\rm xc}$ come across.
Due to its simple analytic structure, this is always the case for the
parametrization of Gross {\em et al.} \cite{GK} The expression of 
Nifos\`\i, Conti and Tosi \cite{conti,nifosi}, however, has to be
handled with some caution for large values of $|\Im \omega|$.


\subsection{A simplified approach for the linewidth of collective modes}

The linewidth of collective electronic modes can be obtained in a
simplified manner \cite{VUC,strips}. 
The idea is to first calculate the modes using
the ALDA. Outside the regime where Landau damping occurs, the modes
will come out undamped, and the mode frequency $\Omega$ is purely real. 
The damping due to dynamical xc effects is then
added on perturbatively. For the one-dimensional case, the resulting 
linewidth (twice the imaginary part of the mode frequency) is given by 
\begin{equation} \label{III.C.1}
\Gamma_p = \frac{|\Re \int dz\: u^*_1(z,\Omega)
\frac{\partial}{\partial z} \sigma_{{\rm xc},zz}(z,\Omega)|}
{\int dz\: n_0(z) |u_1(z,\Omega)|^2} \;.
\end{equation}
From Eq. (\ref{III.B.5}),
the derivative of the stress tensor can be written as
\begin{equation}\label{III.C.2}
\frac{\partial}{\partial z} \sigma_{{\rm xc},zz}(z,\Omega) =
n_0 \frac{\partial}{\partial z} v_{\rm xc,1}\alda (z,\Omega)
- n_0 \frac{\partial}{\partial z} v_{\rm xc,1} (z,\Omega) \;.
\end{equation}
We therefore obtain the linewidth within the DLDA as
\begin{equation} \label{III.C.3}
\Gamma\dlda_p = \frac{ \int dz\: 
\left| \frac{\dy \partial j_1}{\dy \partial z} \right|^2
|\Im f_{\rm xc}^{hom}(n_0,\Omega)|}
{\Omega \int dz \: n_0(z) |u_1(z,\Omega)|^2} \;,
\end{equation}
and within VUC as
\begin{equation} \label{III.C.4}
\Gamma\vuc_p = \frac{ \int dz \: n_0^2(z) 
\left| \frac{\dy \partial u_1}{\dy \partial z} \right|^2
|\Im f_{\rm xc}^{hom}(n_0,\Omega)| }
{\Omega \int dz \: n_0(z) |u_1(z,\Omega)|^2} \;.
\end{equation}
$\Gamma\dlda_p$ depends on the gradient of the current density $j_1$,
whereas $\Gamma\vuc_p$ depends on the gradient of the velocity
field $u_1$. These quantities are related to the density response $n_1$
via the linearized continuity equation
\begin{equation}\label{III.C.5}
i\Omega n_1 = \frac{\partial j_1}{\partial z}
= \frac{\partial}{\partial z}\:(n_0 u_1) \;.
\end{equation}


\section{Results and discussion}

\subsection{Numerical results}

Let us now apply the linear response formalism outlined above  
to the quantum wells whose electronic ground states we have calculated
in Section II. The goal is to study the collective intersubband transitions
between the first two subbands. Our main interest lies in the
damping of these modes.

The standard procedure is to set the external potential $v_{\rm ext,1}$ in
the response equation (\ref{III.A.5}) equal to zero. The resulting 
integral equation for $n_1(z,\omega)$ 
can then be viewed as a complex eigenvalue problem.  The solutions 
with real eigenvalue 1 are the collective modes of the system. 
The real part of the mode frequency, 
$\Omega= \Re \omega$, indicates the position of the mode, and its lifetime
is given by $\tau = \Gamma^{-1}$, where $\Gamma = 2\Im \omega$.

Alternatively, one may consider the photoabsorption
cross section
\begin{equation}\label{IV.1}
\sigma(\omega) = \frac{4\pi\omega}{c} \: \Im \alpha(\omega) \;,
\end{equation}
where $\omega$ is real, and
\begin{equation} \label{IV.2}
\alpha(\omega) = - \frac{2}{E} \int \! dz \: z \, n_1(z,\omega)
\end{equation}
is the dipole polarizability associated with an external potential of the
form $v_1(z,\omega) = E z \cos(\omega t)$. We expect to
find for our quantum wells that $\sigma(\omega)$
is peaked at the frequency of the collective intersubband transition
$\Omega$, with a Lorentzian profile of FWHM $\Gamma$.

We have performed both kinds of calculations for the single and double
quantum well. In addition, we  compare with the linewidths $\Gamma_p$ 
following from the simplified approach presented in Sec. III C.
Results obtained with the various approaches
and approximations are summarized in Tables \ref{table1} and \ref{table2}.
The first columns give the mode frequency $\Omega$ obtained from ALDA
calculations for the collective modes. In both cases, 
$\Omega$ is already quite close to the
experimental result shown in the last column (within 4.2\% for the
single and within 4.9\% for the double square well).
However, as mentioned before, the ALDA does not include damping of the modes.
This means that $\sigma(\omega)$ would consist of a delta peak at $\Omega$.

The second and third columns in Tables \ref{table1} and \ref{table2}
give results for $\Omega$ and $\Gamma$ obtained within the DLDA, using the
parametrizations of Gross/Kohn and Nifos\`\i/Conti/Tosi for $f_{\rm xc}^{hom}$,
respectively. Results within the VUC approximation are given in columns four
and five, again for the two different parametrizations for $f_{\rm xc}^{hom}$.
The first and third lines show $\Omega$ and $\Gamma$ as obtained from a
solution of the response equation for the lowest eigenmode.
In the second and fourth line, we show the position and width of the 
associated peak in the photoabsorption cross section $\sigma(\omega)$.
Finally, the last line gives the linewidths $\Gamma_p$
as obtained from the formulas presented in Sec. III C.

Let us first consider results for the single quantum well, see 
Table \ref{table1}. We note that for each choice of the xc potential
there is an excellent agreement between
the results for $\Omega$ and $\Gamma$ computed with the three different
methods. The results for $\Gamma_p$ are extremely close to the linewidths
obtained using the full calculation for the collective mode or the
photoabsorption cross section, which demonstrates the reliability of the 
linewidth formulas (\ref{III.C.3}) and (\ref{III.C.4}).

We find that the Gross/Kohn parametrization for $f_{\rm xc}^{hom}$
shifts the mode frequency $\Omega$ from its value in ALDA closer towards the
experimental value. The agreement is fairly good within the DLDA, but
the improvement is only small if the VUC approximation is used.
The Nifos\`\i/Conti/Tosi parametrization, on the other hand, leaves the
mode frequency $\Omega$ practically unchanged, within DLDA as well as VUC.

Compared to the experimental value of 0.53 meV, the linewidth $\Gamma$
comes out about 25\%  
too high within DLDA, with only small differences between
the two parametrizations for $f_{\rm xc}^{hom}$ (0.683 and 0.655 meV, 
respectively). By contrast, the VUC linewidth is roughly a factor of
five smaller that the experimental linewidth. It thus seems as if the
DLDA yields a better description of the damping of the modes in the 
square well.

One has to emphasize, however, that the experimental 
linewidth contains not only damping effects due to electron-electron
interaction, but also caused by additional mechanisms such as finite 
temperature, scattering off impurities, and fluctuations of the width of 
the well \cite{estimate}. 
Therefore, our calculations should provide a strictly lower limit
to the measured linewidth. Clearly, this is the case if the VUC approach is 
used, whereas, under this aspect, the overestimation of the electronic
xc damping within the DLDA is substantial.
Note that this effect would be even more drastic in a parabolic well:
in that case, VUC (which satisfies the HPT) would correctly give zero damping,
whereas DLDA would result in an unphysical finite linewidth.

For the double quantum well, however, the situation is different. 
From Table \ref{table2}, we see that the DLDA results are at least
of similar, sometimes even
better quality than for the single well. Again, the mode frequency $\Omega$
agrees better with experiment if the Gross/Kohn parametrization for 
$f_{\rm xc}^{hom}$ is used. In that case, the linewidth is slightly
smaller than in the experiment (1.00 meV compared to 1.17 meV). With the
Nifos\`\i/Conti/Tosi parametrization, it becomes even smaller by 
almost a factor of three (0.403 meV).

The VUC method, on the other hand, clearly fails to provide an adequate
description for the double quantum well. Within the Gross/Kohn parametrization,
the  mode frequency $\Omega$ drastically overshoots the experimental value 
by about 40\%. The opposite happens with the Nifos\`\i/Conti/Tosi 
parametrization, which shifts the ALDA value for $\Omega$ down by more than
1 meV. In both cases, the linewidth is dramatically too high. We also
observe some differences between the results obtained from calculating the
complex mode frequency and the photoabsorption cross section. Moreover,
the linewidth formulas (\ref{III.C.3}) and (\ref{III.C.4}) seem to break down
here. This obvious failure of the VUC method for the double well
calls for some explanation.  We shall therefore examine the nature of the 
electron dynamics in the two quantum wells more closely.


\subsection{Physical picture: collective vs. single-particle}

To get some feeling for the underlying physics of the dynamical processes 
involved, it is helpful to look directly at the local behavior
of various quantities characterizing the response of the electron liquid.

Fig. \ref{fig2} shows the ground-state density $n_0(z)$ for the single
square well and the density profile of its lowest collective mode, $n_1(z)$, 
together with the current density $j_1(z)$ and the velocity field $u_1(z)$.
Calculations have been done within the ALDA. The other approximations 
for the xc potential give very similar results.
All quantities plotted in Fig. \ref{fig2} vary quite smoothly across the
well. Note that for the case of a parabolic well, $n_0(z)$ and $j_1(z)$ 
would have an identical shape, so that the velocity field $u_1(z)$ would 
be uniform.  Here, however, $u_1(z)$ has a shape similar to (though slightly
smoother than) that of $n_0(z)$ and $j_1(z)$, which may be viewed as an 
illustration for the non-parabolicity of the square well.

The same quantities have been plotted for the double quantum well in 
Fig. \ref{fig3}. Here, the ground-state density $n_0(z)$ has two peaks,
the lower one at the narrower well and the higher one at the wider well.
The dip in $n_0(z)$ is at the position of the barrier between the two wells.
The density response $n_1(z)$ does not deviate too much from the one found
for the single well, see Fig. \ref{fig2}, apart for a slight hump at the
barrier. The strongest differences, however, show up in the current density 
and the velocity field. We see from Fig. \ref{fig3} that $j_1(z)$ is very
smoothly varying across the system, apparently ignoring the presence of
the barrier between the wells. The velocity field $u_1(z)$, however, is
strongly peaked at the position of the barrier.

From the examples shown in Figs. \ref{fig2} and 
\ref{fig3}, we see that the most sensitive indicator for the physical nature, 
or rather the degree of ``collectiveness'',  of the motion of the electron 
liquid is the velocity field $u_1(z)$. Two extreme cases have to be 
distinguished: (i) Collective motion, as realized in a perfect way
in a parabolic well. The velocity field is then only slowly varying. 
This is the case on which the HPT is based.
(ii) Single-particle-like motion, occurring for instance
in a tunneling process through
a classically forbidden region, like the barrier between the two wells.
Such a process is characterized by a pronounced peak in the velocity
field, or, more generally speaking, by the presence of large gradients
in $u_1(z)$.  We are thus led to consider more closely the gradients of 
the various quantities involved. 

Note that the domain of validity for the
VUC method is restricted by the conditions \cite{VK1,VK2,VUC} $k\ll
\omega/v_F,k_F$ and $q\ll  \omega/v_F,k_F$, where $k^{-1}$ and $q^{-1}$
are the characteristic length scales for variation of the external
potential and equilibrium density, respectively, and $k_F$ and $v_F$ are
the local Fermi momentum and velocity.
For the problems considered here (calculation of the photoabsorption
cross section), the variation of the external potential is negligible,
i.e. the condition on $k$ is fulfilled. In turn,
a measure for $q$ is given by the quantity $|n_0'(z)|/n_0(z)$.

The top parts of Figs. \ref{fig4} and \ref{fig5} compare the local values
of $k_F$ and $\omega/v_F$ with $|n_0'|/n_0$ for the single and
double well. For the single well, the condition $|n_0'|/n_0 < \omega/v_F,k_F$
is reasonably well satisfied for the central part of the density distribution,
but not towards the borders of the potential well. Note that a similar
behavior would be found for a parabolic well.  For the double well in Fig. 
\ref{fig5}, by contrast, the same condition is much more strongly
violated, due to the fact that there are large density gradients around
the region of the potential barrier in the middle.

From the definition of the velocity (see Eq. (\ref{III.C.5})), we can write
\begin{equation}\label{IV.3}
\frac{n_0'(z)}{n_0(z)} = \frac{j_1'(z)}{j_1(z)} - \frac{u_1'(z)}{u_1(z)} \;.
\end{equation}
The two quantities on the right-hand side are plotted in the lower parts
of Figs. \ref{fig4} and \ref{fig5}, again comparing with the local $k_F$.
We see that for the single quantum well, $|u_1'|/u_1$ lies consistently
below $|j_1'|/j_1$. For the double well, however, the situation is 
opposite: $|u_1'|/u_1$ has sharp structures around the position of the barrier
and is much greater than $k_F$, whereas $|j_1'|/j_1$ is relatively 
smooth over the whole interior of the system.
Note, finally, that for the case of a parabolic quantum well we would
have $n_0'/n_0 = j_1'/j_1$ since here the velocity gradient vanishes.

We thus see immediately from Eqs. (\ref{III.C.3}) and (\ref{III.C.4})
that $\Gamma \vuc \gg \Gamma \dlda$
for the double well is explained by the large values of $|u_1'|/u_1$.
This suggests an additional, more refined local criterion for applicability of 
the VUC approach, requiring that $|\nabla {\bf u}_1({\bf r})|/u_1({\bf r})
\ll |\nabla {\bf j}_1({\bf r})|/j_1({\bf r})$.
This condition is closey related to the hydrodynamical picture inherent in
VUC. It may be viewed as requiring a high degree of collectivity of
the motion of the electron liquid in the system.
This also suggests a pragmatic way to deal with the question of which 
method to choose for approximating the dynamic xc potential for an
arbitrary system: use VUC only
in those regions of space where the electronic motion is predominantly
collective, i.e. where $|\nabla {\bf u}_1({\bf r})|/u_1({\bf r}) <
|\nabla {\bf j}_1({\bf r})|/j_1({\bf r})$. In those regions where the
velocity gradients are large (i.e. the motion is more single-particle like),
use the much more robust DLDA. Note that this prescription automatically
satisfies the HPT.
The so defined VUC-DLDA hybrid scheme yields a linewidth which always
lies below the linewidths of the two ``pure'' schemes, or at most equals
the smaller one.

We have applied the VUC-DLDA hybrid scheme to our quantum wells. 
From Fig. \ref{fig4} it can be seen that for the single well the VUC-DLDA
scheme coincides with the VUC method since here the velocity field is always 
smoother than the current density. In contrast, Fig. \ref{fig5} for the
double well shows that within VUC-DLDA the central part is treated with 
the DLDA, whereas the VUC is used for the outer region. 
Numerical results are shown in Table \ref{table3}.
Compared to DLDA (see Table \ref{table2}),
the resulting mode frequency $\Omega$ is somewhat smaller if the Gross-Kohn
parametrization is used, and almost unchanged with the Nifos\`\i/Conti/Tosi
parametrization. The effect on the linewidth is more pronounced:
for both parametrizations, it is reduced to roughly half of its DLDA value,
and is now substantially smaller than the experimental value, as it should.

Fig. \ref{fig6} illustrates the transition between the ``collective''
and the ``single-particle-like'' regime. We calculate the linewidth and 
lowest collective mode frequency 
(using ALDA and the linewidth formulas of Sec. III C)
for our asymmetric double quantum well with 
variable barrier height, from its maximum value used so far, down to zero
(which then defines a single square well of width 181\AA).
All the while, the electronic sheet density is kept constant.
The calculations have been done using the Gross/Kohn parametrization
(the other parametrization yields similar results).
We see that for decreasing barrier height, the VUC linewidth shrinks much
more rapidly than the DLDA linewidth. The crossover occurs around 
0.3 times the maximum height. This is the region where the mode frequency
becomes comparable to the barrier height. The VUC-DLDA results, on the
other hand, lie consistently below the DLDA results, and merge with 
VUC for small barrier heights, indicating a high degree of collectivity.


\section{Conclusion}   

In this work, we have carried out a comparative study of two approaches
to describe the linear electronic response of quantum wells within a
density-functional framework. The two methods are the dynamical extension
of the ALDA by Gross and Kohn (DLDA) and the recent implementation of 
time-dependent current density functional theory describing dynamical 
exchange and correlation in the language of conventional hydrodynamics (VUC). 
The latter approach is rigorously justified in the limit of slowly varying
density, in the sense that it satisfies the harmonic potential theorem
and other exact Ward identities and symmetries, which in turn are violated
by the DLDA. A third approach, recently proposed by Dobson, B\"unner and
Gross \cite{buenner}, turns out to coincide with VUC in the case of 
one-dimensional inhomogeneity studied here.

Our studies show that the VUC functional should be applied
in general only with a certain caveat, since there are physical situations of
interest where it fails to work. Its domain of applicability was found to 
be closely linked to the structure of the velocity field of the electronic
motion. If the latter is sufficiently smooth - which indicates a high
degree of collective motion - then VUC leads to sensible results.
On the other hand, if there are large gradients of the velocity field 
(as observed around the nonclassical barrier region in the double well),
then VUC is obviously inadequate. By contrast, the DLDA approach was found
to be much less sensitive to the nature of the electronic motion and 
led to a reasonably good description of the double well, although it clearly
overdamps the predominantly collective motion in the single well.

We finally proposed a pragmatic answer to the question of which functional
to choose in a general case. We introduced a hybrid scheme using either 
VUC or DLDA in their respective spatial region of applicability.
This provides a robust method to describe damping of electronic modes of
either collective or single-particle-like nature,
as shown for the case of a double well with variable barrier height.

\acknowledgements

This work was supported by
Research Board Grant RB 96-071 from the University of Missouri and by
NSF grant No. DMR-9706788.  We thank Sergio Conti and Jon Williams for 
useful discussions.



\begin{figure}
\caption{
Full lines: effective Kohn-Sham potential along the growth direction
for the electronic ground state of the single and double square wells.
The electronic sheet densities are $0.97 \times 10^{11}\: \rm cm^{-2}$
and $2 \times 10^{11}\: \rm cm^{-2}$, respectively.
The dashed lines indicate the bound single-particle levels.
The energy scale is such that the Fermi energy lies at zero.
Only the lowest level is occupied in both cases.
The collective transitions we are interested in take place between the
two lowest subbands.
}
\label{fig1}
\end{figure}

\begin{figure}
\caption{
Top: ground-state density profile $n_0(z)$ for the single square well,
see Fig. \ref{fig1}. Middle: density profile $n_1(z)$ of the lowest collective
mode, calculated within the ALDA. Bottom: associated current density
$j_1(z)$ and velocity profile $u_1(z)$. The latter has been scaled with
a factor of 0.05.
}
\label{fig2}
\end{figure}

\begin{figure}
\caption{
Same as Fig. \ref{fig2}, for the double quantum well.
}
\label{fig3}
\end{figure}

\begin{figure}
\caption{
Top: rate of variation of the ground-state density, 
$|n_0'(z)|/n_0(z)$, of the single square well,
versus local $k_F$ and $\omega/v_F$.
Bottom: rate of variation of associated current density $j_1$ and velocity
profile $u_1$.
}
\label{fig4}
\end{figure}

\begin{figure}
\caption{
Same as Fig. \ref{fig4}, for the double quantum well.
}
\label{fig5}
\end{figure}

\begin{figure}
\caption{
Ratio of linewidth $\Gamma_p$ and lowest collective mode frequency $\Omega$
for the double well with barrier height varying from zero to one,
where one corresponds to the full height.
Calculations are done with ALDA and the linewidth formulas of Sec. III C,
using the VUC, DLDA and VUC-DLDA hybrid scheme, as indicated.
}
\label{fig6}
\end{figure}

\begin{table}
\caption{ Lowest collective intersubband transition of the single
GaAs/AlGaAs square well, calculated with various approaches for the
linearized xc potential. The superscript 1 (2)
stands for the parametrization by Gross/Kohn (Nifos\`\i/Conti/Tosi)
for $f_{\rm xc}^{hom}$. $\Omega$ is the frequency of the mode,
$\Gamma$ its width. The subscripts indicate the computational scheme: search 
for the eigenmode of the system as pole in the complex $\omega$-plane ($m$), 
calculation of the photoabsorption cross section ($\sigma$),
and using the simplified approach from Sec. III C ($p$).
All numbers are given in meV.
}
\label{table1}
\begin{tabular}{ccccccc}
                  &  ALDA  & DLDA$^1$ & DLDA$^2$ & VUC$^1$ & VUC$^2$ 
& Exp. \\ \hline
$\Omega_m$  &  10.25 & 10.63 & 10.23 & 10.31 & 10.24 & 10.7 \\
$\Omega_\sigma$   &     & 10.63 & 10.24 & 10.31 & 10.24 &      \\
$\Gamma_m$   &        & 0.683 & 0.655 & 0.128 & 0.104 & 0.53 \\
$\Gamma_\sigma$  &     & 0.683 & 0.639 & 0.128 & 0.104 &      \\
$\Gamma_p$   &        & 0.686 & 0.636 & 0.130 & 0.104 &       \\
\end{tabular}
\end{table}

\begin{table}
\caption{ Same as Table \ref{table1}, for the double square well.
}
\label{table2}
\begin{tabular}{ccccccc}
                  &  ALDA  & DLDA$^1$ & DLDA$^2$ & VUC$^1$ & VUC$^2$ 
& Exp. \\ \hline
$\Omega_m$  &  13.85 & 14.24 & 13.88 & 20.64 & 12.55 & 14.6  \\
$\Omega_\sigma$   &       & 14.24 & 13.89 & 20.23 & 12.89 &       \\
$\Gamma_m$   &        & 1.00  & 0.403 & 8.55  & 4.15 & 1.17  \\
$\Gamma_\sigma$   &       & 1.00  & 0.401 & 8.52  & 3.52 &       \\
$\Gamma_p$   &        & 0.988 & 0.406 & 11.07 & 6.50 &       \\
\end{tabular}
\end{table}

\begin{table}
\caption{ Same as Table \ref{table2}, for the double square well, using the
VUC-DLDA scheme.
}
\label{table3}
\begin{tabular}{cccc}
                  &  VUC-DLDA$^1$ & VUC-DLDA$^2$ &  Exp. \\ \hline
$\Omega_m$ &   14.07 & 13.87 &  14.6  \\
$\Gamma_m$ &    0.620 & 0.194 &  1.17  \\
$\Gamma_p$         & 0.605 & 0.192 &       \\
\end{tabular}
\end{table}

\newpage
\Large
\begin{center} Fig. 1 \end{center}
\unitlength1.0cm
\begin{picture}(15.0,19.0)
\put(-10.0,-16.0){\makebox(15.0,19.0){
\includegraphics{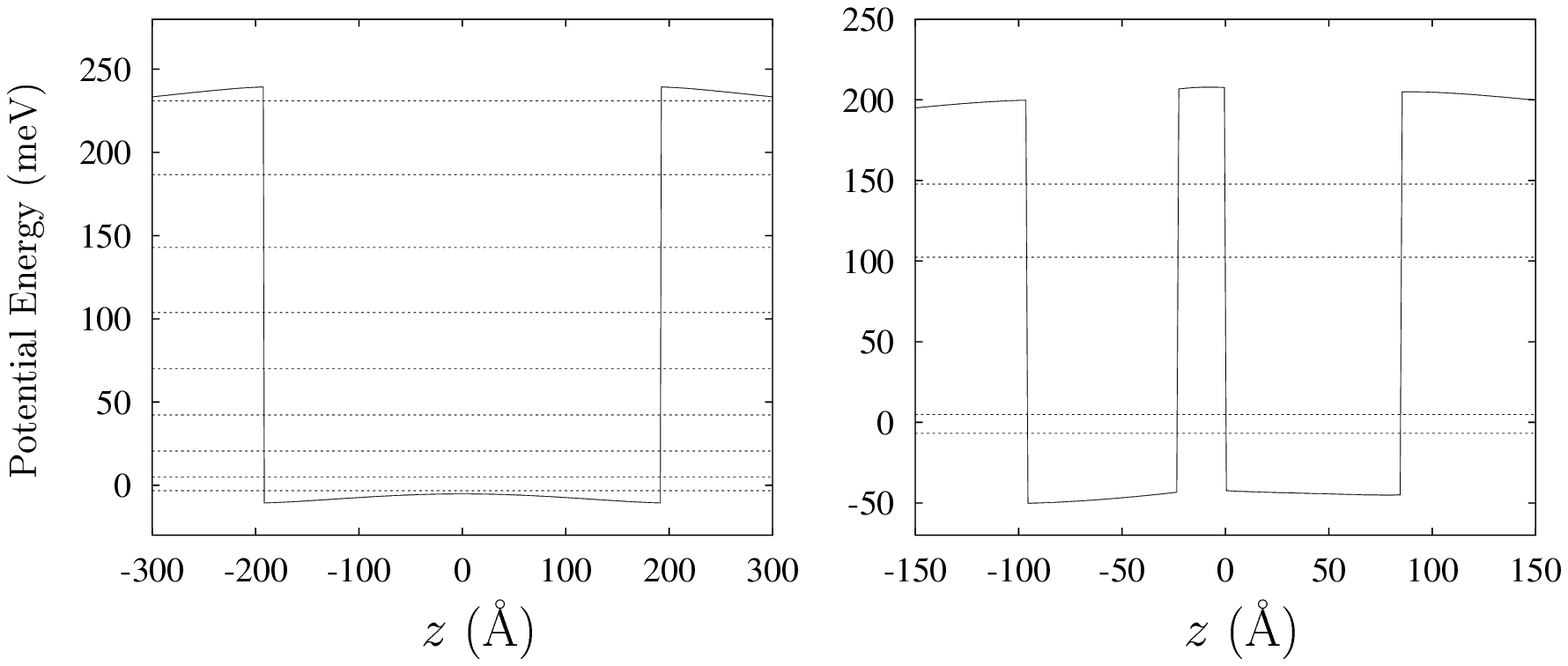}
}}
\end{picture}
\newpage 
\begin{center} Fig. 2 \end{center}
\begin{picture}(15.0,19.0)
\put(-11.0,-15.5){\makebox(15.0,19.0){
\includegraphics{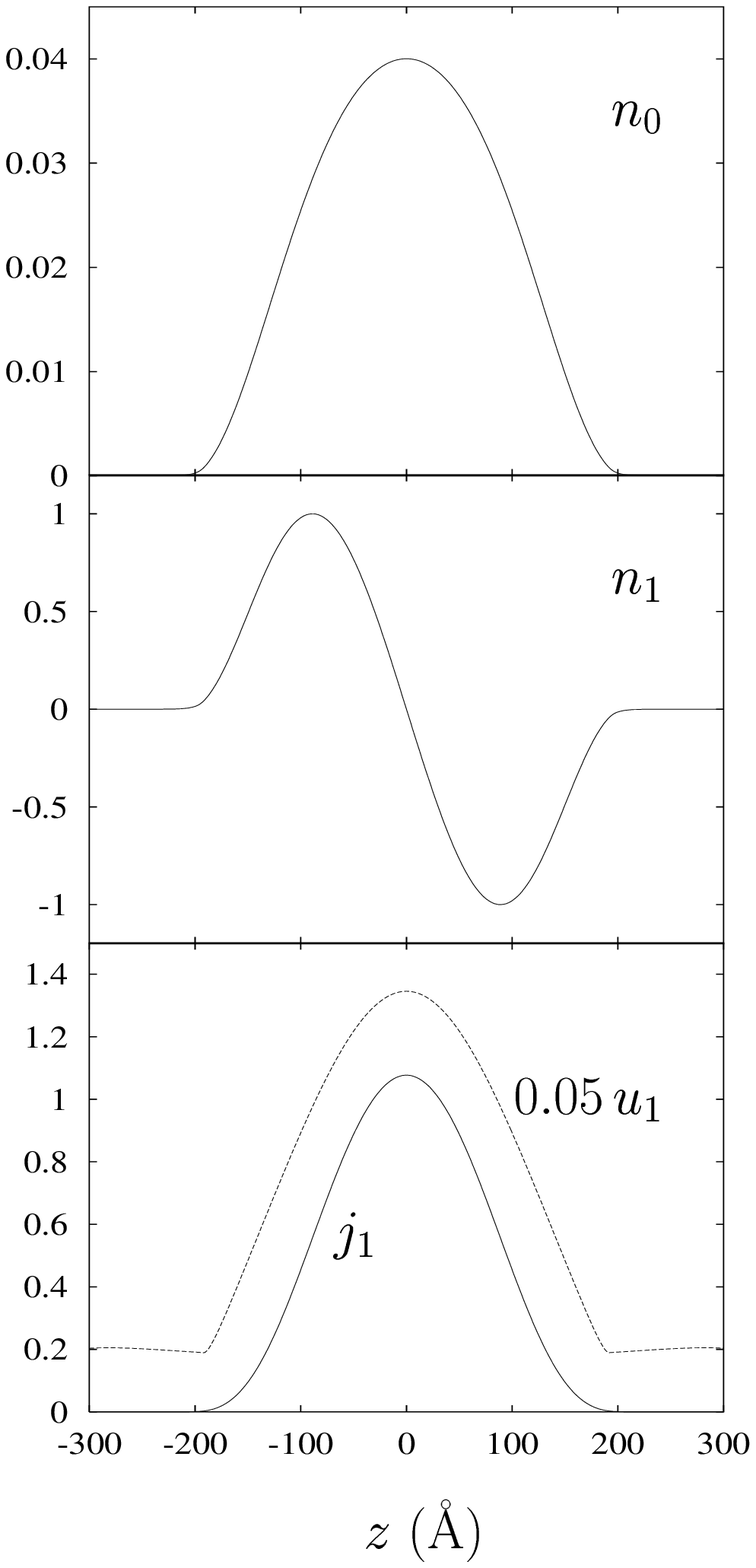}
}}
\end{picture}
\newpage 
\begin{center} Fig. 3 \end{center}
\begin{picture}(15.0,19.0)
\put(-11.0,-15.5){\makebox(15.0,19.0){
\includegraphics{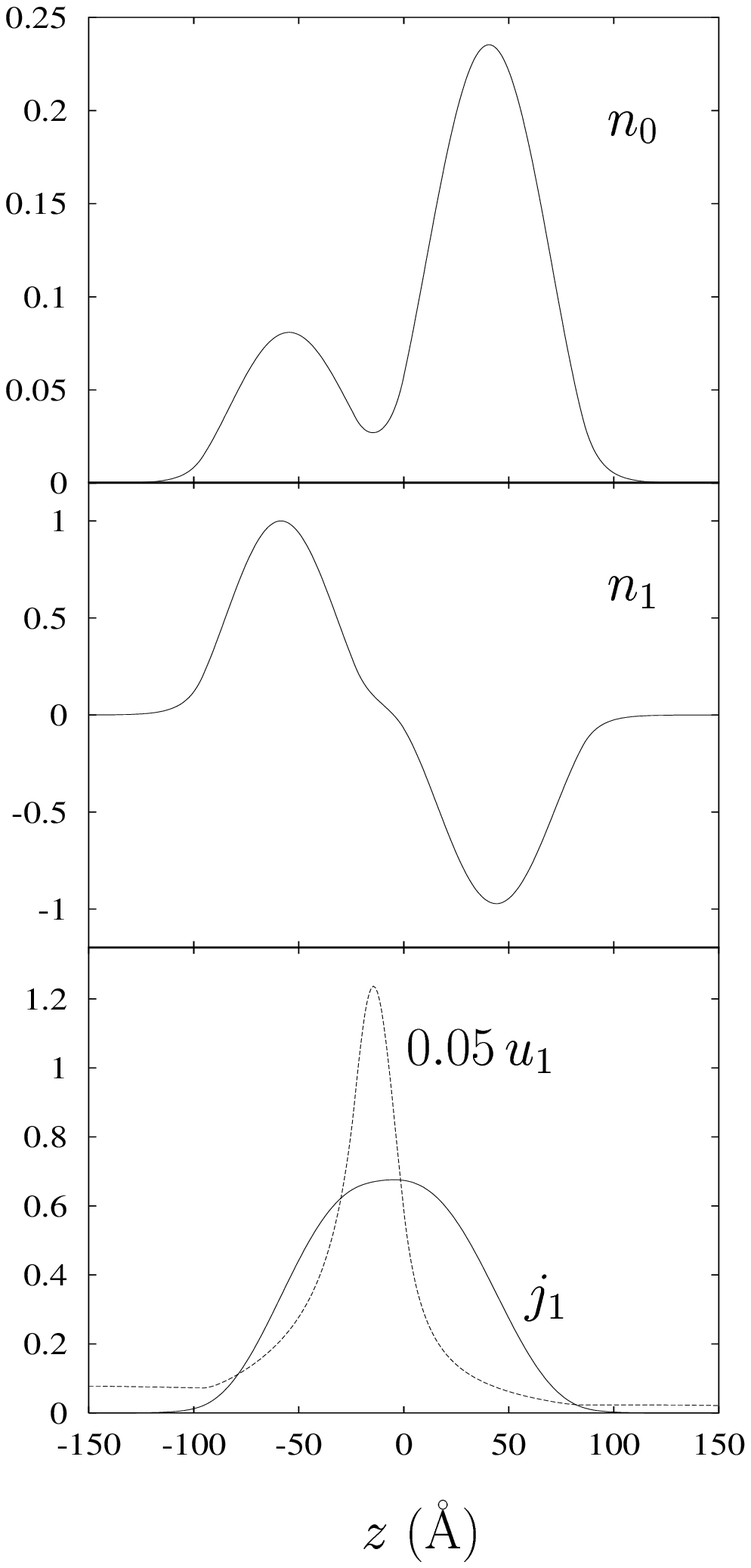}
}}
\end{picture}
\newpage 
\begin{center} Fig. 4 \end{center}
\begin{picture}(15.0,19.0)
\put(-11.0,-16.0){\makebox(15.0,19.0){
\includegraphics{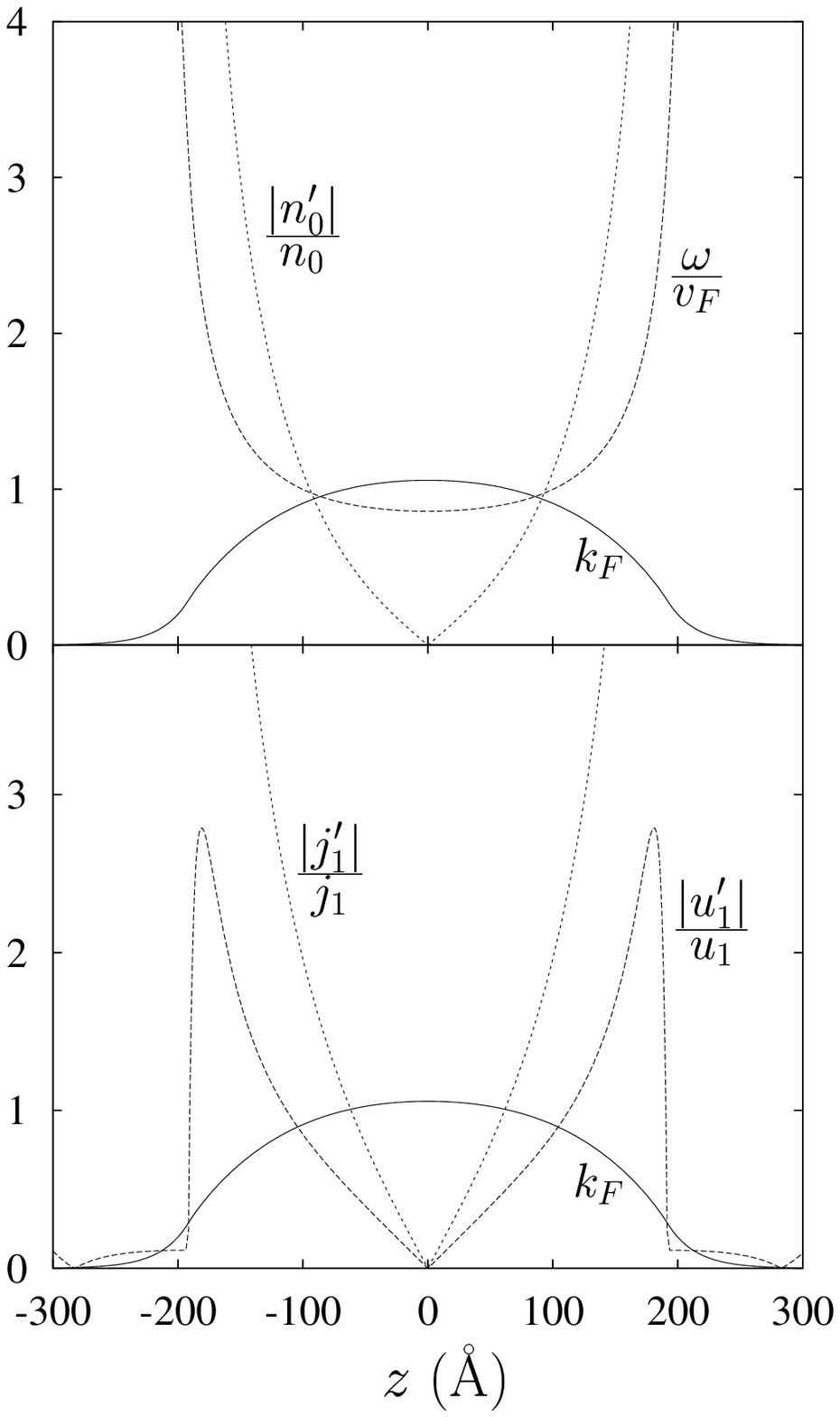}
}}
\end{picture}
\newpage 
\begin{center} Fig. 5 \end{center}
\begin{picture}(15.0,19.0)
\put(-11.0,-16.0){\makebox(15.0,19.0){
\includegraphics{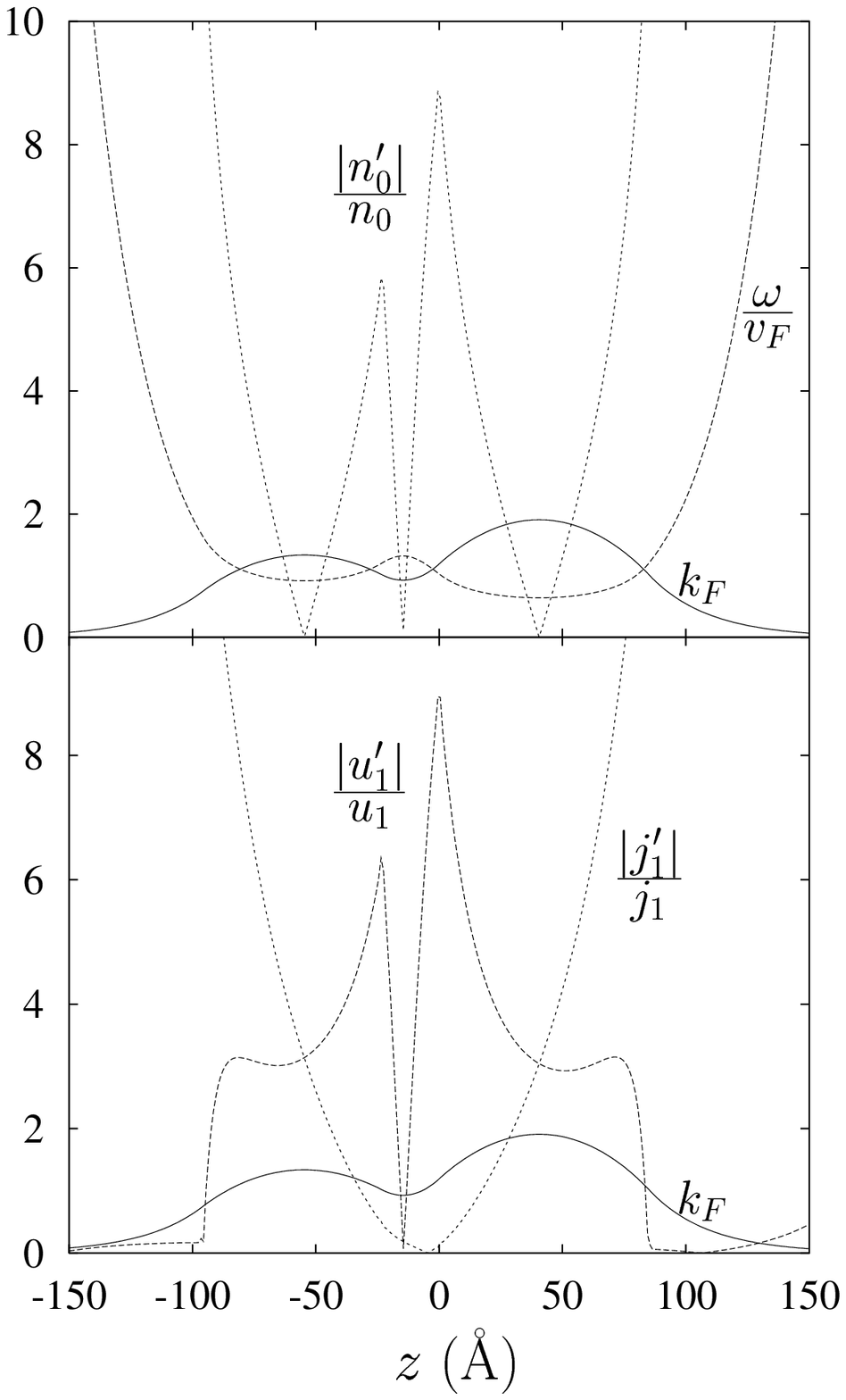}
}}
\end{picture}
\newpage 
\begin{center} Fig. 6 \end{center}
\begin{picture}(15.0,19.0)
\put(-11.0,-16.0){\makebox(15.0,19.0){
\includegraphics{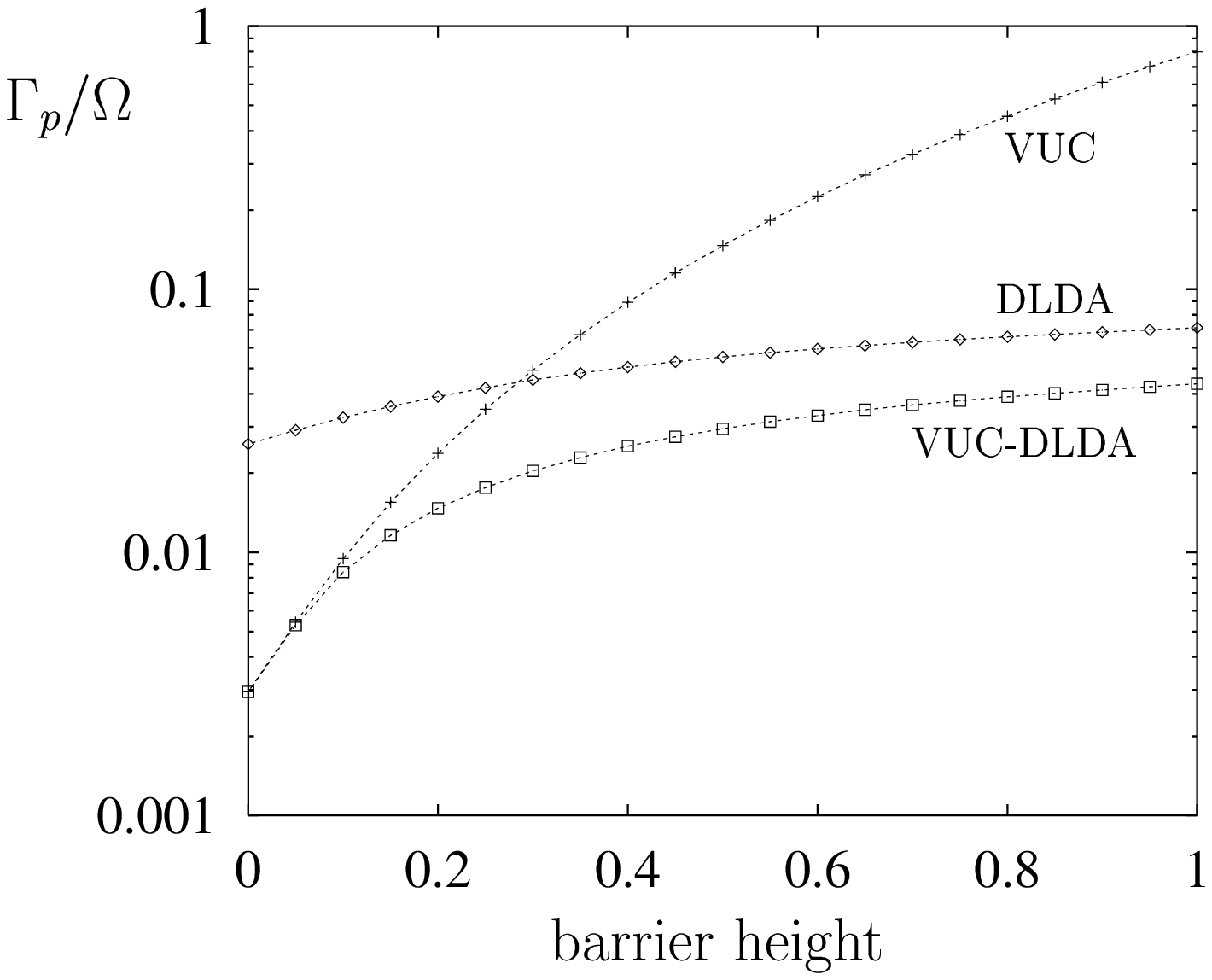}
}}
\end{picture}

\begin{references}
\bibitem{estimate}
To estimate the contribution of the last effect to 
the line broadening, note that changing the width 
of the well by one monolayer (about 3\AA) shifts the real part of the
mode frequency by about 0.1 meV.
\end{references}
\end{document}